\documentclass[dvips]{article}

\usepackage{icrc2011}
\usepackage[hyphens]{url}
\usepackage{hyperref}
\usepackage{breakurl}

\title{The status of gamma-ray astronomy}
\newcommand{\etal}{\MakeLowercase{\textit{et al. }}} % "et al."
\shorttitle{Funk, S. \etal The Status of gamma-ray astronomy}

\authors{Stefan Funk$^{1}$ }
\afiliations{$^1$Kavli Institute for Particle Astrophysics and Cosmology, Department of Physics
and SLAC National Accelerator Laboratory, Stanford University, Stanford, California 94305, USA}
\email{funk@slac.stanford.edu}

\abstract{Gamma-ray studies are an essential tool in our search for
  the origin of cosmic rays. Instruments like the Fermi-LAT, H.E.S.S.,
  MAGIC and VERITAS have revolutionized our understanding of the high
  energy Universe. This paper describes the status of the
  very rich field of gamma-ray astrophysics that contains a wealth of
  data on Galactic and extragalactic particle accelerators. It is the
  write-up of a rapporteur talk given at the 32nd ICRC in Beijing,
  China in which new results were presented with an emphasis on the
  cosmic-ray related studies of the Universe.}
\keywords{ }

\begin{document}
\maketitle

%Begin the section.
\section{Introduction and General statements}

This paper is intended to give a summary of the results reported from
the sessions OG 2.1-2.5 of the 32$^{\mathrm{nd}}$ ICRC that took place
in August of 2011 in Beijing, China. The basis of this write-up is the
Rapporteur talk on these sessions (available at
{\url{http://indico.ihep.ac.cn/getFile.py/access?contribId=1388&sessionId=16&resId=0&materialId=slides&confId=1628}})
The aim of this paper is to give an overview over the vibrant field of
gamma-ray astronomy at the time of writing (focusing on results
presented at the ICRC but mentioning where necessary the bigger
picture). Where appropriate I will use updated plots if journal
publications have become available since the ICRC. The sessions in OG
2 cover topics related to the origin of cosmic rays (CRs) as probed by
both X-ray and gamma-ray measurements. In particular, gamma-ray
studies that address fundamental physics questions, such as the
particle nature of dark matter are not part of OG2 (they are presented
in HE3) and will thus not be part of this summary. Traditionally, only
very few X-ray results are presented at the ICRC and in the following
I will solely focus on the gamma-ray results. A total of 101 talks and
105 posters were presented. These were split into the different
sessions as follows:

\begin{figure*}[!t]
  \centering
  \includegraphics[width=\textwidth]{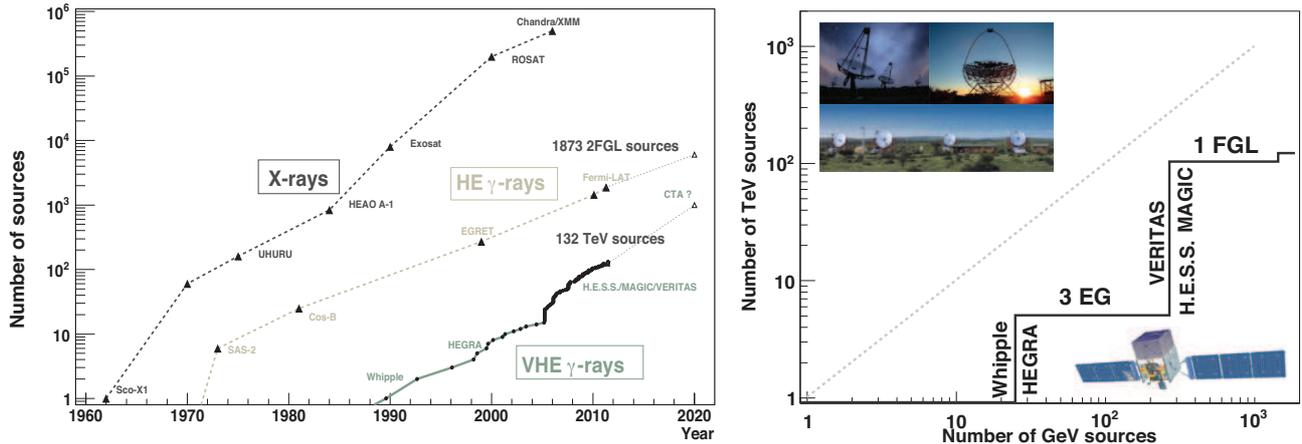}
  \caption{{\bf{Left:}} {\emph{Kifune}} plot (named after T. Kifune,
    who first showed a similar plot at the 1995 ICRC in Rome), showing
    the number of sources detected over time for various
    wavebands. {\bf{Right:}} Comparison between the number of TeV
    sources and the number of GeV sources.}
  \label{kifune}
\end{figure*}

\begin{itemize}
\item OG 2.1: Diffuse X-ray and gamma-ray emission: 11
\item OG 2.2: Galactic sources: 60
\item OG 2.3: Extragalactic sources: 46
\item OG 2.4: Gamma-ray bursts: 11
\item OG 2.5: Instrumentation: 65
\end{itemize}

Taking a step back and looking at the overall situation of the field
since the last ICRC two years ago, the field has clearly seen several
important milestones. 

%  \begin{figure}[!t]
%   \vspace{5mm}
%   \centering
%   \includegraphics[width=2.in]{icrc2011_fig01.eps}
%   \caption{Simple figure example}
%   \label{simp_fig}
%  \end{figure}

\begin{figure*}[th]
   \centering
   \includegraphics[width=0.9\textwidth]{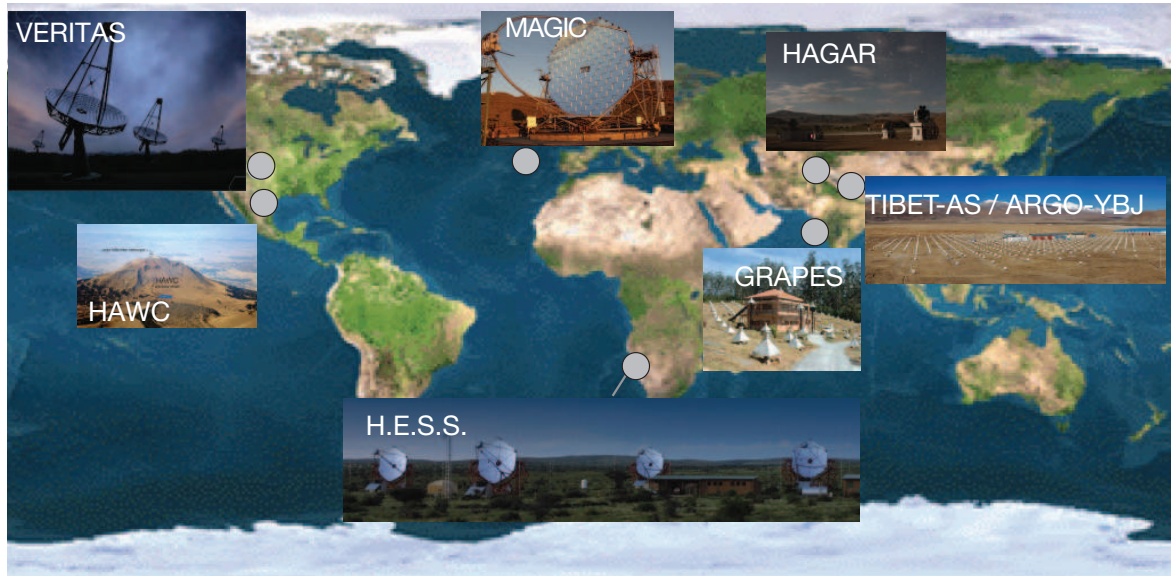}
   \caption{The ground-based gamma-ray world as of spring 2012,
     showing currently operating gamma-ray instruments that presented
     results at the 32nd ICRC in Beijing. }
   \label{gamma_world}
  \end{figure*}

\begin{itemize}
\item The {\emph{Fermi}}-LAT team has released their first
  (1FGL)~\cite{1FGL} and second (2FGL)~\cite{2FGL} catalogs of the GeV
  sky. The 2FGL catalog contains 1873 sources and has important
  implications for the origin of cosmic rays in its own right but is
  also extremely important for the ground-based instruments that
  measure at higher energies to guide them towards interesting
  objects. The idenfication of the GeV sources is a very active field
  of study. In 2FGL 127 out of the 1873 2FGL sources are firmly
  identified (based on criteria such as correlated periodicity,
  correlated variability, or matching spatial morphology). Among those
  127, 83 are pulsars, 28 are Active Galactic Nuclei (AGN), 6 are
  supernova remnants (SNRs), 4 are High-mass Binaries (HMB), 3 are
  Pulsar Wind Nebulae (PWNe), 2 are normal galaxies, and one is a
  nova. A larger number of 2FGL sources have positional associations
  (though no firm identifications) but 572 of the 2FGL sources are not
  even positionally associated with any object in the catalogs that
  have been searched. Given the very large number of unidentified
  sources interesting (and unexpected) physics might be waiting to be
  discovered.
\item {\emph{MAGIC-II}} has started operation by commissioning the
  second 17-m telescope and has reported first results at this
  ICRC~\cite{CortinaMAGIC, ZaninICRC}. The energy threshold of the
  stereo system is 50 GeV for a Crab-like
  spectrum~\cite{ZaninICRC}. In addition, the two other large
  ground-based arrays {\emph{VERITAS}} and {\emph{H.E.S.S.}} continue
  to produce first-class science~\cite{HolderICRC, EmmaICRC} and are
  in the process of upgrading their systems which will improve
  sensitivity and lower the energy threshold in both cases.
\item {\emph{HAWC}}, a large water Cherenkov array in the Sierra Negra
  in Mexico~\cite{GoodmanICRC}, has started operation as a prototype
  array called {\emph{VAMOS}} (with 7 out of the 300 tanks). A first
  skymap collected in 24 hours of lifetime with 4 of the
  {\emph{VAMOS}} tanks was shown, including 16.6 million events with a
  threshold of 15 hit PMTs~\cite{BaughmanICRC}. By the spring of 2012
  the HAWC collaboration expects to have deployed 30 tanks which would
  give a sensitivity comparable to {\emph{MILAGRO}}.
\item The field is vibrant and lots of projects are ongoing (see
  Figure~\ref{gamma_world}, showing the ``ground-based
  gamma-ray world''). There seems to be a solidification of the techniques
  that are used in gamma-ray astronomy - we have learned how to build
  the best instruments. The main future projects use imaging
  atmospheric Cherenkov telescopes (as in the case of {\emph{CTA}}) or
  a large array of water tanks (as in the case of {\emph{HAWC}}) or a
  combination of them with added particle detector arrays (as in the
  case of {\emph{LHAASO}}). These two main techniques are very
  complimentary to each other in terms of energy range, sky coverage
  and angular resolution and promise to provide exciting science well
  into the next decade. It should also be mentioned that two of the
  pioneering observatories of the field have stopped operation since
  the last ICRC: the {\emph{CANGAROO-III}} array and the
  {\emph{Whipple}} 10m telescope, the grand-father of all modern
  imaging Cherenkov telescopes~\footnote{W. Benbow quoted T. Weekes
    during the ICRC on the Whipple 10m telescope as saying "Thank god
    for Mkn421!"}).
%\item Lastly, it is my sad duty to remember two of the main
%  contributors to our field over the past decades who passed away
 % during the last two years: Okkie de Jager and Simon Swordy. 
\end{itemize}

The success of the field can be illustrated by the number of sources
detected at GeV and at TeV energies (see Figure~\ref{kifune}). At the
time of this writing 1873 GeV and 132 TeV sources are known and we
have a good knowledge about the objects that dominate the GeV and the
TeV sky.  However, more importantly, the field is moving beyond
quantity towards a qualitatively better understanding about the
high-energy processes at work in these sources as will hopefully
become apparent in this write-up. In the following, I will give my
personal selection of highlights. 
%As usual, I apologize in advance to
%everyone whose work I have either unfairly omitted or whose work I
%have inadvertently misrepresented.

\section{Diffuse emission Studies}
The study of diffuse emission at the ICRC has focused mostly on
the Galactic diffuse emission (as opposed to the isotropic
extragalactic diffuse emission). The Galactic diffuse emission is
produced by the interaction of CRs (protons, nuclei and electrons)
with interstellar gas and radiation fields~\cite{FermiEGB}. While
diffuse emission is the dominant source of astrophysical photons at
GeV energies ($\sim$ 80\% of all LAT photons are Galactic diffuse
emission), at TeV energies this emission is sub-dominant, due to the
rapidly falling spectrum compared to most galactic gamma-ray sources.
Eleven contributions have been submitted to {\emph{OG 2.1 Diffuse
    emission studies}} (Fermi-LAT: 3, TIBET: 2, ARGO-YBJ: 1, INTEGRAL:
1, MILAGRO: 1, Interpretation: 3).

  \begin{figure*}[th]
   \centering
   \includegraphics[width=0.8\textwidth]{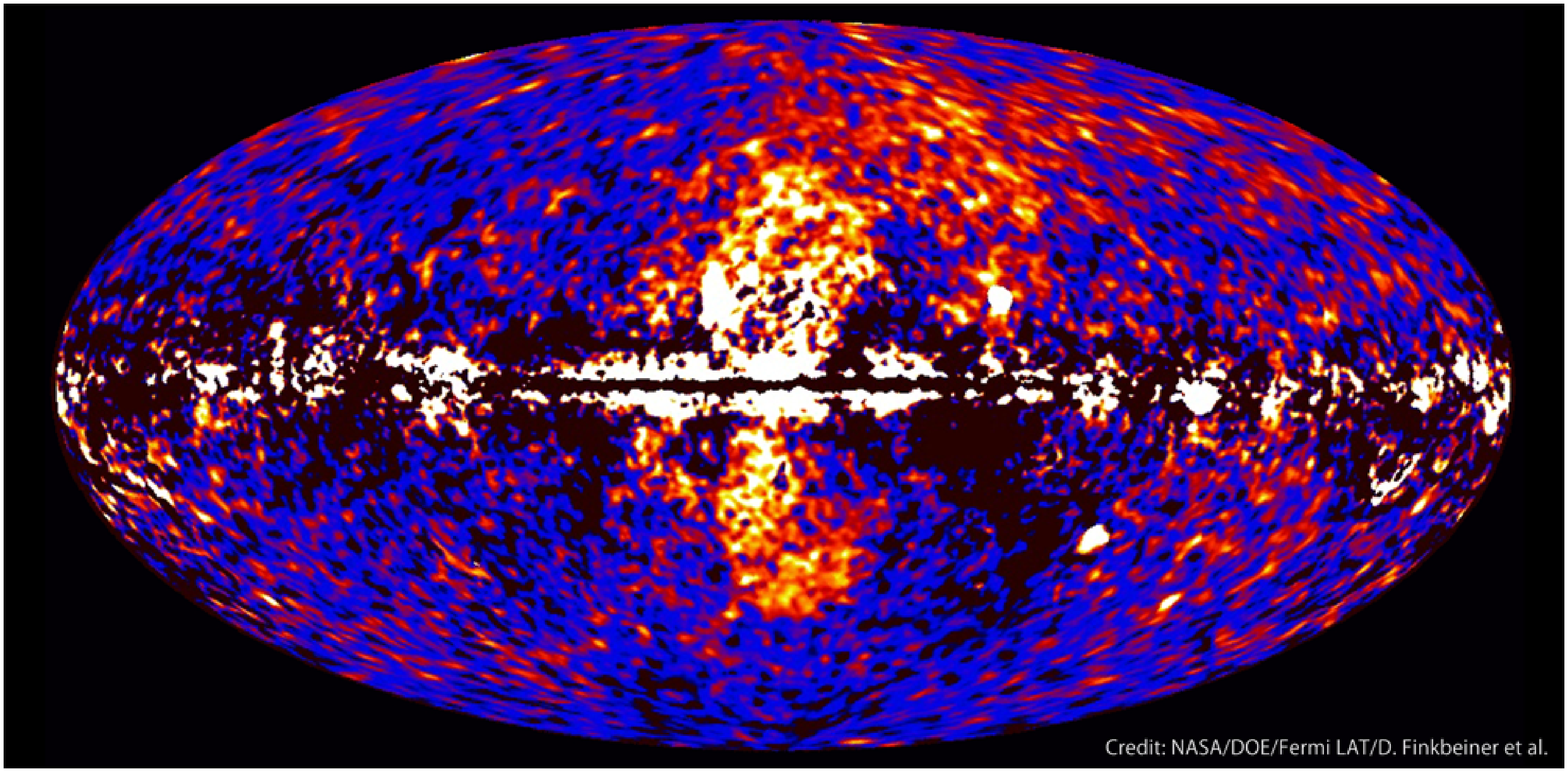}
   \caption{Fermi-LAT all-sky residual image after subtracting a
     diffuse model using photons in the range between 1 GeV and 10
     GeV. Reproduced from nasa.gov, contributions by~\cite{SuICRC,
       MertschICRC, ChernyshovICRC}. Credit: NASA/DOE/Fermi LAT/D. Finkbeiner
     et al.  
     }
   \label{fig:fermibubbles}
  \end{figure*}

  The study of the diffuse emission has benefited tremendously from
  the launch of the Fermi-LAT representing a significantly improved
  instrument compared to previous ones in this energy range. The main
  result since the last ICRC (at which a thorough discussion on the
  absence of the EGRET excess was presented) came through the
  detection of the so-called {\emph{Fermi bubbles}}~\cite{SuICRC} (see
  Figure~\ref{fig:fermibubbles}). These residual (after removing the
  Galactic diffuse, and the bright sources) large-scale structures
  (scale height $\sim 10$kpc if at the distance of the Galactic
  Center) are located towards the Galactic center and extend to large
  scale heights. Their main characteristics are not yet firmly
  established (a publication by the LAT team is still pending), but
  they seem to show rather hard energy spectra (compared to the
  Galactic diffuse). Whether they have sharp edges and whether they
  are symmetric with respect to the plane is more difficult to
  establish and seems to depend on the choice of the Galactic diffuse
  model that is removed~\cite{FermiDiffuse2}. If they are symmetric
  about the Galactic center, they are likely related to some past
  activity of the super-massive black hole in the center of our
  Galaxy.

Two contributions were presented to explain the properties of the
Fermi bubbles ~\cite{ChernyshovICRC, MertschICRC} that both explain
them as Inverse Compton emission of accelerated electrons. An
alternative model invoking pion decay of accelerated protons has been
published elsewhere~\cite{CrockerAharonian}. The general problem these
models have to face are: a) the cooling time scales if assuming that
the bubbles are generated by Inverse-Compton scattering accelerated
electrons, b) the fact that there is no apparent limb-brightening in
the emission but rather a flat-top profile and c) the total power in
the emission at a distance of 10~kpc from the super-massive black
hole. ~\cite{ChernyshovICRC} explain the bubbles as generated
by accelerated electrons through first-order Fermi-acceleration. The
shocks are generated by the tidal disruption of stars by the central black
hole which generated a system of hundreds of concentric shock fronts
that accelerate electrons in the bubbles. These shocks are then
thought to be constricted in the Galactic disk to generate the
apparent pinched bubble shape. ~\cite{MertschICRC} suggest that the
bubbles are generated by accelerated electrons through second-order
Fermi-acceleration in plasma wave turbulence in the interior of the
bubbles. Whether these explanation or the alternative hadronic
scenario is correct is (at least) partially testable with Fermi-LAT
data and it should be expected that by the time of the next ICRC we
will have improved our understanding of these fascinating structures.

The Galactic diffuse emission is by itself a useful tool in studying
Cosmic rays in our Galaxy. ~\cite{MizunoICRC} presented an analysis of
the Fermi-LAT observations of diffuse emission towards the outer
Galaxy. This region has the advantage of not being susceptible to the
distance ambiguity in the velocity separation of the gas and is
suitable to study CRs associated with both arms and inter-arm
regions. Two important conclusions were drawn from these studies: a)
the shape of the emissivity spectra does not change significantly with
Galactocentric distance but agree well with the model for the locally
measured CRs. b) the Cosmic ray density (i.e.\ the emissivity
normalization) is larger than expected in the outer regions of the
Galaxy, pointing to a rather large CR halo (larger than 10 kpc) or a
flatter CR source distribution than the canonical pulsars or supernova
remnants. ~\cite{MaICRC} summarized the results of the studies of the
diffuse emission at TeV energies with ARGO-YBJ (a resistive place
chamber array that has an energy threshold of $\sim$300 GeV). ARGO-YBJ
detected diffuse TeV gamma-ray emission from two regions along the
galactic plane described by a power-law with spectral index $\Gamma =
2.9 \pm 0.3$. The normalization is significantly larger than the
expectation from the extrapolation of the Fermi-LAT diffuse emission,
so unresolved sources might contribute to this emission in a
significant way. 

The Cygnus region also received considerable attention. It is the
brightest region in gamma rays in the northern sky from GeV to
multi-TeV energies and updates were presented by the Fermi-LAT,
MILAGRO and VERITAS collaborations. \cite{BonamenteICRC} showed that
the TeV emission seen by MILAGRO is dominated by two sources both of
which seem to coincide with Fermi-LAT sources and which exhibit energy
spectra up to $\sim 100$TeV. The higher-angular-resolution VERITAS
array has performed a survey of the Cygnus region in
2007-2009~\cite{AliuICRC}. The observations resolve one of the two
MILAGRO sources (MGRO J2019+37) into a very complex region consisting
of at least one point-source coincident with the SNR CTB 87A, and an
additional large-scale structure that coincides with the brightest
part of the MILAGRO emission. At energies above 10 GeV, the Fermi-LAT
also resolves a very complex structure in the Cygnus region consisting
of diffuse emission and individual point
sources~\cite{TibaldoICRC}. One of the sources that can be detected
when removing the galactic diffuse emission is the SNR
$\gamma$-Cygni. When also removing this bright point-source at
energies $> 10$GeV, a residual diffuse emission of size $\sim
10^{\circ}$ can be detected coinciding with a region bounded by
ionized gas (so-called {\emph{Cocoon}}) that shows a rather hard
energy spectrum. The interpretation put forward is that this feature
shows freshly accelerated particles (combination of electrons and
protons) that leave the $\gamma$-Cygni SNR and stream into a
low-density cavity (see also~\cite{FermiCocoon}).

In summary, all these studies show that there is a large amount of
information on Galactic cosmic rays that can be drawn from the study
of the Galactic diffuse emission. The Fermi bubbles are residual
structures that are clearly seen beyond a reasonable range of Galactic
diffuse models. Their explanation is not straightforward but
different models should be testable with LAT data in the future.

\section{Galactic Sources (OG 2.2)}

One of the main motivations for the study of Galactic gamma-ray sources
has direct relevance for one of the main topics of this conference:
the search for the origin of cosmic rays. However, as we have learned
by now, the process is far from straightforward, given that competing
processes emit gamma rays in the relevant energy range and one of the
main challenges is to distinguish gamma rays emitted through hadronic
processes ($\pi^0$-decay) from those originating in leptonic processes
(inverse Compton scattering and bremsstrahlung). Multi-wavelength
observations are one crucial ingredient in the quest for separating
the two and therefore the identification of Galactic gamma-ray
emitters with astrophysical objects known at other wave bands is an
important prerequisite in the study of the origin of cosmic
rays. Sixty contributions have been submitted to {\emph{OG 2.2
    Galactic Sources}} (H.E.S.S.: 16, MAGIC: 7, Fermi-LAT: 6, VERITAS:
5, ARGO-YBJ: 4, AGILE: 1, NCT: 1, GRAPES-3: 1, HAGAR: 1, Shalon: 2,
CTA: 1, Interpretation: 11, Methods: 3).

The last decade has shown that gamma-ray emission is prevalent and
occurs in many different kinds of sources. By now, several classes of
objects are known to emit gamma rays in the GeV and TeV band in our
Galaxy: supernova remnants (SNRs), and Pulsar Wind Nebulae (PWNe) are
by far the most abundant ones as can be seen from table~\ref{tab:1},
which lists the firmly identified TeV objects. In addition, pulsars
are prevalent at GeV energies in the Fermi-LAT sky and the most
energetic Galactic one - the Crab - is now also seen at higher
energies ($>$100 GeV) with VERITAS~\cite{CrabVERITAS} and
MAGIC~\cite{CrabMAGIC, CrabMAGIC2}. In addition, a handful of
gamma-ray binaries are now detected, mostly both at GeV and TeV
energies and in addition, one nova has been observed by the Fermi-LAT
team at GeV energies~\cite{FermiNova}. Also at GeV energies, globular
clusters are clearly detected, emitting gamma rays most probably
through the combined emission of their population of milisecond
pulsars (MSPs). This ICRC has seen a report of a possible detection of
the globular cluster Terzan~5 with H.E.S.S.\ at higher
energies~\cite{DomainkoICRC}. The slight offset of the gamma-ray
emission from the center of globular cluster and the rather large extension
of the emission is challenging for the Globular cluster
interpretation. The chance coincidence is however quoted at the level
of $10^{-4}$.

%In addition, there are lots of unidentified sources, both at GeV and
%TeV energies within our Galaxy, the most obvious of which is the
%Galactic center source which could be associated with the black hole
%in the center of our Galaxy, or with more {\emph{standard}}
%astrophysical sources such as the supernova remnant Sgr~A~East or a
%pulsar wind nebula.

\begin{table*}
\begin{center}
\begin{tabular}{c | c | c || c | c}
Object & Discovered & Year & Type & Method\\
\hline
Crab Nebula & Whipple & 1989 & PWN & Position\\
Crab Pulsar & MAGIC & 2008 & Pulsar & Periodicity\\
RX\,J1713.7$-$3946 & CANGAROO & 2000 & SNR & Morphology\\
Cassiopeia A & HEGRA & 2001 & SNR & Position\\
RX\,J0852.0$-$4622 & CANGAROO & 2005 & SNR & Morphology\\
G0.9+0.1 & H.E.S.S. & 2005 & SNR & Position\\
HESS\,J1825$-$137 & H.E.S.S. & 2005 & PWN & ED Morphology\\
MSH 15$-$52 & H.E.S.S. & 2005 & PWN & Morphology\\
LS 5039 & H.E.S.S. & 2005 & $\gamma$-ray binary & Periodicity\\
HESS\,J1303$-$631 & H.E.S.S. & 2005 & PWN & ED Morphology\\
PSR\,B1259$-$63 & H.E.S.S. & 2005 & $\gamma$-ray binary & Variability\\
Vela~X & H.E.S.S. & 2006 & PWN & Morphology\\
LS I+61 303 & MAGIC & 2006 & $\gamma$-ray binary & Periodicity\\
Kookaburra (Rabbit) & H.E.S.S. & 2006 & PWN & Morphology\\
Kookaburra (Wings) & H.E.S.S. & 2006 & PWN & Morphology\\
HESS\,J0632$+$057 & H.E.S.S. & 2007 & $\gamma$-ray binary &
Periodicity\\
HESS\,J1731$-$347 & H.E.S.S. & 2007 & SNR & Morphology\\
RCW 86 & H.E.S.S. & 2008 & SNR & Morphology\\
SN 1006 & H.E.S.S. & 2009 & SNR & Morphology\\
Tycho & VERITAS & 2010 & SNR & Position\\
\hline
\end{tabular}
\caption{Firmly identified Galactic TeV objects.}\label{tab:1}
\end{center}
\end{table*}

The H.E.S.S.\ Galactic plane survey, for a long time a source of new
discoveries has been expanded since the last ICRC and now comprises
2300 hours of data~\cite{GastICRC}. By now, over 60 sources of VHE gamma
rays have been found within its current range of $l = 250^{\circ}$ to
$60^{\circ}$ in longitude and $|b| leq 3.5^{\circ}$ in latitude and the
sensitivity of the survey is below 2\% of the Crab flux everywhere
within the survey region. Interestingly, nearly a third of the sources
detected in this region remain unidentified or confused.

\subsection{The Crab}
At the time of the last ICRC there was no reason to believe that we had not firmly
understood the Crab Pulsar and Nebula. The Crab Pulsar is the most
energetic pulsar in our Galaxy. The pulsed emission was thought to be
generated by curvature radiation in the outer magnetosphere, following
the detailed measurements of the Crab pulsar with the Fermi-LAT. The
Crab nebula, the brightest steady TeV gamma-ray source, was often used
by ground-based instruments as a test-source which was observed during
the commissioning of a new instrument due to its flux stability. The
emission of the Crab nebula is comprised of synchrotron and inverse
Compton emission from relativistic electrons that were thought to be accelerated
somewhere in the vicinity of the termination shock. The fact that the
synchrotron emission extends to $\sim 100$ MeV was taken as a sign
that the Crab Nebula could in fact accelerate particles to energies
close to 1 PeV assuming that the acceleration happened in a region in
which the magnetic field was close to the average field of $\sim 300
\mu$G for the nebula.

  \begin{figure}[!t]
   \vspace{5mm}
   \centering
   \includegraphics[width=0.5\textwidth]{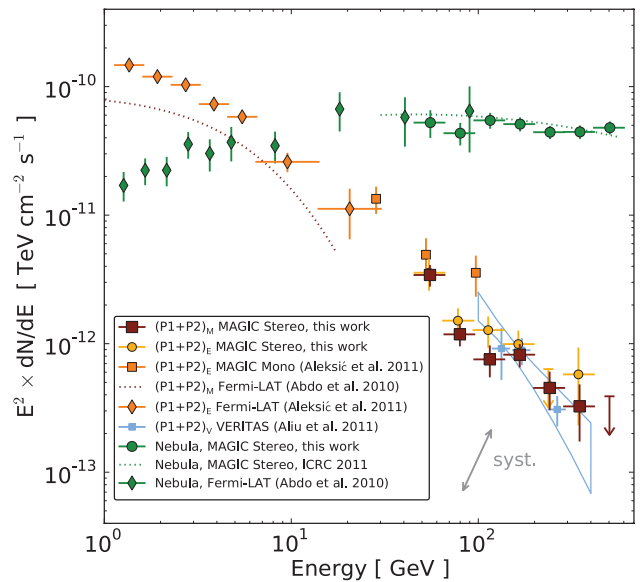}
   \caption{Compilation of emission spectra from the Crab
     Pulsar (P1+P2) as measured by the Fermi-LAT~\cite{FermiCrab},
     VERITAS~\cite{CrabVERITAS} and MAGIC~\cite{CrabMAGIC2}. Reproduced
   from~\cite{CrabMAGIC2}. }
   \label{fig::CrabPulsar}
  \end{figure}

  However, around the time of the ICRC several exciting new
  observations demonstrated that some of the old assumptions about the
  Crab pulsar and nebula were wrong: {\emph{The Crab Nebula Flares
      detected by AGILE and Fermi-LAT}}~\cite{FermiCrabFlares,
    FermiCrabFlares2} and the detection of {\emph{pulsed emission from the
      Crab Pulsar up to beyond 100 GeV}}~\cite{CrabVERITAS,
    CrabMAGIC2}.

  The overall spectral energy distribution (SED) of the Crab Nebula is
  well sampled from radio all the way to TeV gamma rays. The Fermi-LAT
  is observing in an interesting energy range, covering the band where
  the synchrotron emission drops sharply and the inverse Compton
  emission starts to rise (which happens at $\sim 100$ MeV. The TeV
  instruments have traditionally observed beyond the peak of the
  inverse Compton emission. MAGIC~\cite{ZaninICRC} reported their
  first stereo data, taken with MAGIC-II for which the first flux
  point is at 58 GeV, clearly the lowest energy threshold of all the
  operating IACTs. A joint MAGIC and Fermi-LAT data fit, constraining
  the peak of the Inverse Compton emission to $59 \pm
  6_{\mathrm{stat}}$ GeV, demonstrates that the combination of
  ground-based and space-based instruments have the capability to
  severely constrain the properties of the Crab Nebula. It should be
  noted that this ICRC also saw the first detection of the Crab Nebula
  with a compton telescope. ~\cite{HuangICRC} reported on a possible
  $4\sigma$-detection with the NCT (Nuclear Compton Telescope) in the
  range between 0.2 and 2 MeV.

  The first clear indication that something was not well understood
  came from simultaneous Fermi-LAT and AGILE detections of flaring
  episodes from the Crab Nebula in the highest-energy part of the
  synchrotron emission at energies around 100 MeV. In particular the
  Fermi-LAT observed a "superflare" in April 2011, where the flux in
  the synchrotron component increased by a factor of $\sim 30$ over
  the average emission. The doubling time was $\sim 8$
  hours. Light-crossing time arguments point to an extremely compact
  region for the emission of the
  flares~\cite{FermiCrabFlares2}. Studying the spectral evolution of
  the synchrotron component during the flares suggests that the flares
  are caused by relativistic beaming of a small-scale region within or
  close to the termination shock of the PWN. An alternative
  explanation was presented by~\cite{YuanICRC} which employs a
  phenomenological model in which the gamma-ray flares are produced in
  large ``knots'' that are known from X-ray observations.
  Simultaneous high-angular resolution X-ray Chandra observations did
  not reveal any regions of significant variation in the nebula during
  the gamma-ray flares. At the ICRC, {\emph{ARGO-YBJ}} reported a
  $\sim 3 \sigma$ enhancement of the Crab flux during the LAT flare at
  a median energy of 1 TeV~\cite{VernettoICRC}. While not expected in
  the current models, it would be extremely interesting if those
  synchrotron flares could also be seen in the Inverse Compton
  component (it is expected at significantly higher energies than 1
  TeV). However, such a flux enhancement of the IC component was not
  confirmed by the much more sensitive MAGIC~\cite{ZaninICRC} or
  VERITAS array~\cite{HolderICRC}, although their observations were
  not strictly simultaneous with the LAT flares. Also, non-detection of
  the 2007 AGILE-detected flare was reported with
  MILAGRO~\cite{BraunICRC}.

  In addition to the exciting new observations on the Crab Nebula, the
  Crab Pulsar also received a lot of attention (see
  Figure~\ref{fig::CrabPulsar} through the fact that both
  VERITAS~\cite{McCannICRC} and MAGIC~\cite{SaitoICRC} report pulsed
  emission up to $\sim 400$ GeV. HAGAR~\cite{SinghICRC} reported upper
  limits at energies $> 250$GeV, albeit not at the sensitivity level
  of VERITAS and MAGIC. The spectrum of the Crab pulsar can be well
  fitted with a broken power law (as opposed to an exponential power
  law), clearly indicating that curvature radiation is not a likely
  mechanism at least for the high-energy emission (and possibly even
  for the whole emission from the Crab, given the smooth transition
  between the GeV and TeV spectrum). One possible radiation mechanism
  for the highest energy pulses is Inverse Compton scattering (e.g.\
  from secondary pairs). Another interesting property is that the
  pulses seem significantly narrower than those measured by the
  Fermi-LAT, and that in addition the ratio of the amplitude of the
  two pulses changes with increasing energies. One possible
  explanation of this observed narrowing is that the region where
  acceleration occurs shrinks towards the neutron star with increasing
  energy. ~\cite{OtteICRC} presented an interesting possibility of
  doing Lorentz Invariance tests with pulsars, based on an idea
  by~\cite{Kaaret1999}, by searching for a shift in the peak position
  of the Crab pulsar between the Fermi-LAT and VERITAS. While these
  limits currently are not competitive with the AGN and GRB
  observations done by the Fermi-LAT and IACTs, they do have the
  advantage of being able to more easily distinguish between
  source-intrinsic and propagation effects and might be exploited
  competitively with a future instrument such as CTA.

\subsection{Other PWNe}
PWNe are still the dominant Galactic population of identified objects
at TeV energies. The efficiency of converting spin-down power into
gamma rays is typically in the range of 1-10\% and a wide range of
pulsar ages is seen. Interesting news in this area is that there is a
growing number of objects where the pulsar has been detected with
Fermi-LAT (often through a blind search procedure,
e.g.~\cite{FermiBlind}) and the PWN has then subsequently been
detected with ground-based TeV instruments. Examples of this are the
PWN in CTA~1, detected by both Fermi-LAT and VERITAS~\cite{AliuICRC},
and the gamma-ray emission in the complicated field of SNR G284.3-1.8
that was detected with H.E.S.S.\ and might comprise a PWN and a
point-like source coinciding with the new Fermi-LAT detected
$\gamma$-ray binary 1FGL\,J1018.6$-$5856~\cite{EmmaICRC2}.

\subsection{Shell-type supernova remnants}
The observations of shell-type supernova remnants is directly related
to the origin of cosmic rays, given that these are the prime candidate
sources for the acceleration of Galactic cosmic rays to the energy of
the knee. While still no smoking gun (unambiguous) proof has been
found to relate shell-type SNRs with the origin of cosmic rays,
considerable progress has been made since the last ICRC, mainly
through the release of the 1FGL and 2FGL catalogs. 2FGL lists 58
Fermi-LAT gamma-ray sources spatially coincident with SNRs and PWNe
including such important objects as Cassiopeia A, Tycho's SNR, the
Cygnus Loop, W51C, W44 and IC443, and the TeV-bright SNRs
RX\,J1713.7$-$3946 and RX\,J0852.0$-$4622 (Vela Junior). Before
discussing individual objects in more detail, it is worth noting that
combining measurements from Fermi-LAT and TeV instruments yields
energy spectra over 5-6 orders of magnitude in energy

  \begin{figure}[!t]
   \vspace{5mm}
   \centering
   \includegraphics[width=0.5\textwidth]{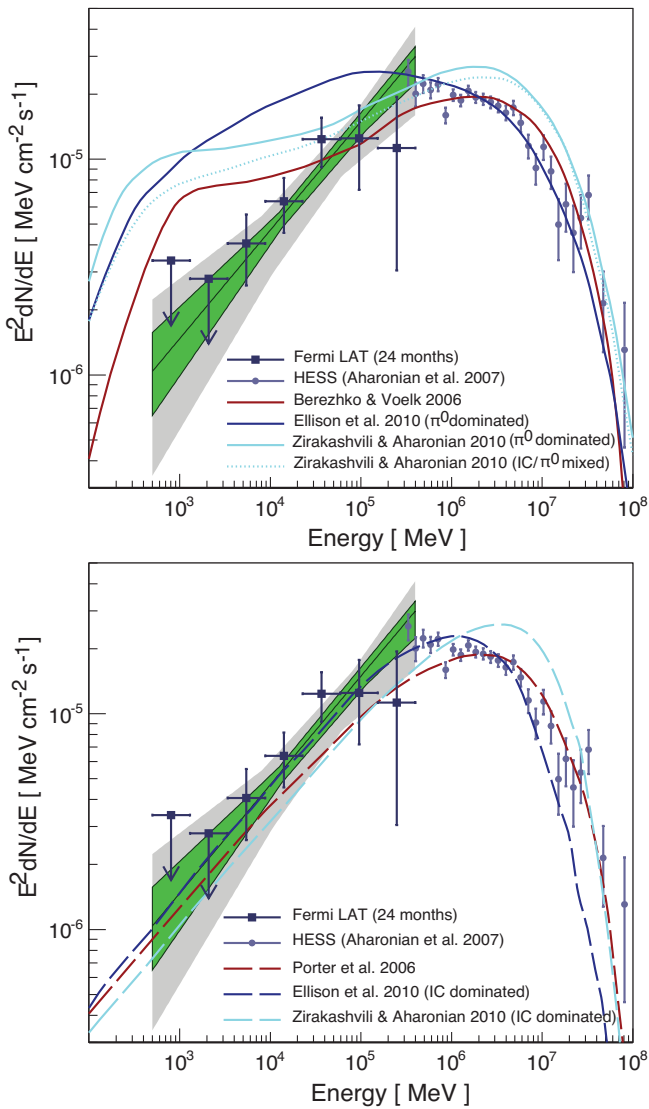}
   \caption{Gamma-ray spectra of RX\,J1713.7$-$3946. Shown is the
     Fermi-LAT spectrum ~\cite{Fermi1713} and the H.E.S.S.-measured
     spectrum~\cite{HESS1713} with pre-Fermi-LAT-measurement hadronic
     and leptonic model curves (for details
     see~\cite{Fermi1713}). Upper limits are set at 95\% confidence
     level. Reproduced from~\cite{Fermi1713}. The gamma-ray
     measurements suggest a leptonic origin of the emission, although
     hadronic models with a very hard proton spectrum can not be ruled
     out.}
   \label{fig::1713}
  \end{figure}

  Possibly the most prominent of these cases is the young ($\sim 2000$
  years) SNR RX\,J1713.7$-$3946 which was discussed by Aharonian in
  his summary talk~\cite{AharonianICRC} (see also
  Figure~\ref{fig::1713}). This object has long been the best
  candidate for TeV gamma-ray emission stemming from hadronic
  interactions, although this claim has been extremely
  controversial. The Fermi-LAT measurements for this object resemble
  more what had been expected before those measurements in the case of
  a leptonic scenario~\cite{Fermi1713} (making the case that protons
  might still be accelerated in the remnant but might not have enough
  ambient target density to interact and produce a flux of gamma rays
  that is larger than the leptonic ``guaranteed'' channel). However,
  Aharonian reminded that {\emph{``life might be more complicated''}}
  and that the proton spectrum in a close-by molecular cloud might be
  extremely hard ($\Gamma=1.5$) and the resulting gamma-ray spectrum
  could therefore fit the Fermi-LAT and H.E.S.S.\ data
  well. ~\cite{WeinsteinICRC} presented the combined Fermi-LAT and
  VERITAS data for the young Tycho SNR in which again through SED
  modelling the leptonic model seems strongly disfavored. No cutoff is
  found in the VERITAS data and by equating the acceleration time to
  the SNR age, the maximum proton energy obtained is $> 300$ TeV
  suggesting acceleration to close to the knee (see
  also~\cite{MorlinoCaprioli}). However, Tycho's SNR is a very weak
  source both in the Fermi-LAT data and for VERITAS and one should be
  careful in drawing strong conclusions from these faint
  sources. While the verdict on these cases might still be open, this
  should serve as a reminder that fitting the spectral energy
  distribution alone will render it extremely difficult to distinguish
  between hadronic and leptonic scenarios. A more robust smoking gun
  feature might be needed to settle this question in the future. One
  class of objects where this might ultimately be possible in the
  future are the GeV-bright mid-aged SNRs interacting with molecular
  clouds such as W51C or W44. The Fermi-LAT might ultimately be able
  to detect the pion-cutoff feature below 100 MeV. However, it should
  be stressed that such analyses will be very complicated due to the
  combination of the bright Galactic diffuse emission and the rapidly
  changing effective area at these energies.

  \begin{figure}[!b]
   \vspace{5mm}
   \centering
   \includegraphics[width=0.5\textwidth]{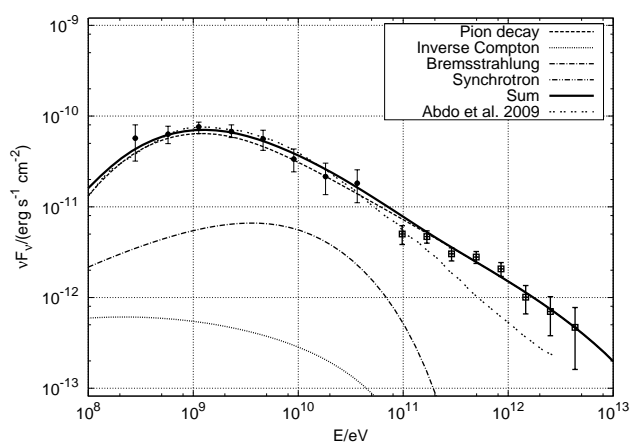}
   \caption{Gamma-ray spectrum of W51C, a mid-aged remnant interacting
     with molecular cloud detected with the Fermi-LAT~\cite{FermiW51C}
     and MAGIC~\cite{MAGICW51C}. Shown are hadronic model curves, one
     from the Fermi-LAT publication that match the Fermi-LAT
     data~\cite{FermiW51C} and one with a somewhat harder proton
     spectrum that match the combination of Fermi-LAT and MAGIC
     data~\cite{MAGICW51C}. Reproduced from~\cite{MAGICW51C}. }
   \label{fig::W51C}
  \end{figure}

  ~\cite{CarmonaICRC} presented a study of W51C at TeV energies with
  the MAGIC telescope (see Figure~\ref{fig::W51C}). The bulk of the
  TeV gamma-ray emission coincides with shocked molecular material and
  suggests an extrapolation of the Fermi-LAT spectrum. Interesting in
  this regard is that MILAGRO detected a faint source at a median
  energy of 35~TeV coinciding with W51C. If this is correct, then a
  second spectral component might be emerging, something that will be
  tested with HAWC and CTA in the future.

This ICRC also saw the convergence of three of the eminences in our
field: Drury, Aharonian and V{\"o}lk were present and thus is was an
appropriate question when~\cite{BochowICRC} asked whether these three
(also known as DAV) were right in their seminal paper in
1994~\cite{DAV} in which they predicted the gamma-ray flux from SNRs
if these were the sources of hadronic cosmic rays. ~\cite{BochowICRC} used
the H.E.S.S.\ Galactic plane survey in conjunction with Green's
catalog of 274 radio SNRs and determined upper limits to the gamma-ray
emission from SNRs that are not detected. The results indicate that
the upper limits are in the right ballpark (acceleration efficiency
around 10\%), although significantly lower than that in some cases. As
Drury put it at the conference: {\emph{``DAV survived a falsification
    test. That doesn't prove it, but it makes it a little more
    plausible.''}}. 

Further results were reported on deeper observations of Vela Junior
with H.E.S.S.\ which suggests a very good agreement between the TeV
gamma rays and the non-thermal X-rays from this
object~\cite{PazArribasICRC}. \cite{BoyerICRC} reported on the discovery
of $^{26}$Al in Vela Junior with COMPTEL and ~\cite{WeinsteinICRC}
showed results that demonstrate the detection of parts of
$\gamma$-Cygni with the VERITAS array. Finally, an object that was not
detected is SN~1987A in the LMC. ~\cite{BerezhkoICRC} presented the
expected gamma-ray flux from SN1987A. Shortly afterwards
~\cite{KominICRC} presented deep H.E.S.S.\ observations that put upper
limits below these expectations. This by itself it turned out was not
a contradiction, because the H.E.S.S.\ data were accumulated
constantly over the last 6 years during which time the predicted flux
significantly increased. Thus the direct comparison between the
average flux upper limits over the last 6 years with the prediction
for the gamma-ray flux of SN 1987A as it is today is somewhat
misleading. The results show, however, that we might be very close to
detecting this object. It would be extremely exciting to see this
object evolve in gamma rays!

\subsection{Binary Systems}
Binary systems are objects that contain a neutron star or a black hole
and a companion star. While relatively few binaries have been
discovered at GeV and TeV energies, these objects are important
testbeds of our understanding of particle acceleration in
astrophysical objects. The periodic occurrence of the same environmental
conditions for the accelerator makes these objects one of the closest
things a gamma-ray astronomer can get to a physical
{\emph{``experiment''}} and can help to distinguish between external
properties and the properties of the accelerator itself.  Before the
ICRC four gamma-ray binaries had been established at TeV energies
(LS\,5039, LS\,I+61\,303, PSR\,B1259$-$63 and possibly Cyg X-1).
The picture might be similar to what has been described before for the
Crab Nebula: the more data we collect, the more detailed and rich the
systems appear. One example is that the light curves of several of the
binaries show very complex behavior with long-term variability
overlaid on top of the periodicity as e.g.\ demonstrated in
LS\,I+61\,303~\cite{JoglerICRC}

\begin{figure}[!t]
   \vspace{5mm}
   \centering
   \includegraphics[width=0.5\textwidth]{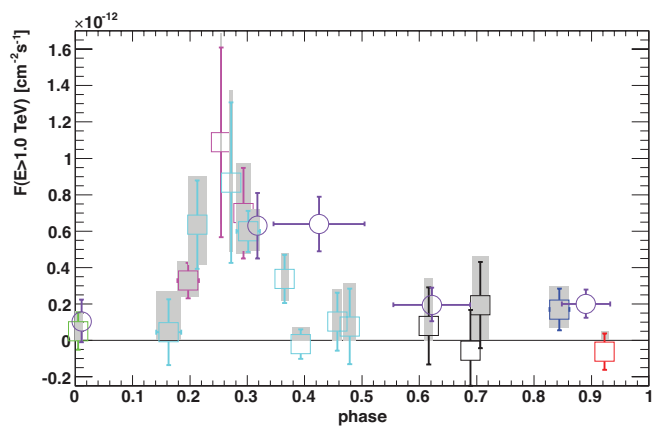}
   \caption{Folded gamma-ray light curve for HESS\,J0632+057, assuming
     a period of 321 days~\cite{Bongiorno2011} as detected by
     SWIFT. Shown are H.E.S.S. (circular markers) and VERITAS (open
     square) measurements. The colors of the markers indicate the
     different periods of the
     observations. See~\cite{MaierSkiltonICRC} for details. Reproduced
     from~\cite{MaierSkiltonICRC}.}
   \label{fig::0632}
 \end{figure}

 In addition, two objects have recently been confirmed to be binary
 objects through the periodic behavior of the emission:
 HESS\,J0632+057 at TeV energies and 1FGL\,J1018.6$-$5856 at GeV
 energies. HESS\,J0632+057 was first detected in the H.E.S.S.\
 Galactic plane survey~\cite{HESSScanII} and was unusual in being one
 of the very few point-like sources detected in the original
 survey. The multi-frequency data led ~\cite{Hinton2009} to suggest
 that this is a binary system, in spite of the fact that no
 periodicity had been detected at any wave band. Long-term SWIFT
 monitoring in X-rays established the periodicity of $321 \pm 5$ days
 recently~\cite{Bongiorno2011}, strongly suggesting a binary nature of
 the underlying object and thus making it the first gamma-ray binary
 detected at gamma-ray energies first. The combination of VERITAS and
 H.E.S.S.\ data folded with the orbital period detected by SWIFT shows
 first hints of a periodicity in the TeV data as
 well~\cite{MaierSkiltonICRC} (see Figure \ref{fig::0632}. This source
 is not detected at GeV energies with Fermi-LAT
 yet. 1FGL\,J1018.6$-$5856 might be another object of this class. It
 was detected with Fermi-LAT and subsequently found to exhibit
 periodic emission at a period of 16.6 days. H.E.S.S.\ also detects
 TeV emission from this region but the picture is more complex at
 higher energies~\cite{EmmaICRC2}. H.E.S.S.\ detects a point-like
 source coinciding with 1FGL\,J1018.6$-$5856 on top of an extended
 structure which might be unassociated with the binary object. Finally,
 when discussing these systems, the most massive binary star system
 in our Galaxy, Eta Carinae, should be mentioned. Particle acceleration
 has been predicted in systems like Eta Carinae through shocks in the
 wind-wind interaction zone. The Fermi-LAT clearly detected Eta
 Carinae during periastron. Also, there seems to be relatively clear
 evidence for two spectral components in the GeV emission, suggesting
 two populations of particles. The low-energy component (measured from
 0.5-8 GeV) shows a cutoff and is stable over time, while the
 high-energy component (above 10 GeV) is clearly varying with
 time~\cite{WalterICRC}. The source of these two components is
 controversial. It could be either electrons responsible for the
 low-energy component through Inverse Compton scattering and protons
 responsible for the high-energy component through
 pion-decay~\cite{WalterICRC}, or two populations of electrons
 generated by the double shock structure of the wind-wind interaction
 zone~\cite{BednarekICRC}. It should be added that the 50-hour
 sensitivity curves for H.E.S.S.\ touches the extrapolation of the
 Fermi-LAT high-energy component, so this is certainly an interesting
 target also for TeV instruments in the future.

Before turning the attention to ``real'' extragalactic sources one
detection should be mentioned: in deep observations of the LMC (90.4
hours) the H.E.S.S.\ collaboration~\cite{KominICRC} has detected a source
coincident with the most energetic pulsar known (PSR\,J0537$-$6910)
that is also known to have an X-ray pulsar wind nebula (detected with
Chandra). The spectrum of the emission is rather soft ($\Gamma = 2.7
\pm 0.2$) but given the energetics, the amount of gamma-ray flux seems
plausible ($\epsilon = 0.08$ where epsilon is the amount of spin-down
power from the pulsar turned into gamma-ray emission). This source is
therefore very likely a gamma-ray PWN and as such is the first of
these objects detected outside our Galaxy. No detection of diffuse
emission from this object at TeV energies has been reported. 

\section{Extragalactic Sources (OG 2.3)}

\begin{figure*}[!t]
  \centering
   \includegraphics[width=0.8\textwidth]{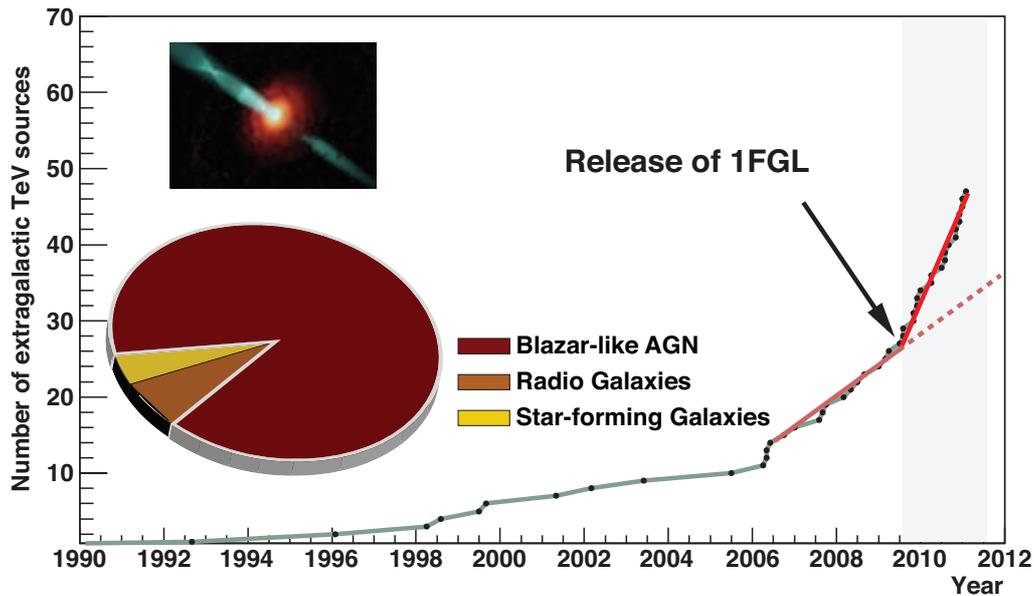}
   \caption{Number of extragalactic TeV gamma-ray sources as a
     function of time. At the time of writing 48 sources are
     known, 43 blazar-like objects, 3 radio galaxies and 2
     star-forming galaxies. Clearly an acceleration of the detection
     rate can be seen that coincides with the release of the first
     Fermi-LAT catalog (1FGL). Inset: Credit to Cosmovision, a group
     led by Dr. Wolfgang Steffen of the Instituto de Astronomia, UNAM,
     Ensenada, Mexico. }
   \label{fig::KifuneExtragalactic}
\end{figure*}

The study of the extragalactic sky at gamma-ray energies has received
a tremendous boost through the launch of the Fermi-LAT (see
e.g. Figure~\ref{fig::KifuneExtragalactic}). Beyond the wealth of
information contained in the GeV data itself the Fermi-LAT catalogs
coupled with the the all-sky monitoring of time-varying processes
provides a very important guidance for the the scheduling of
observations with the rather narrow-field-of-view ground-based TeV
instruments. It should be stressed that while contemporaneous
observations of the Fermi-LAT with HAWC and with CTA would be of great
complementary value, the 10-year mission for the LAT is far from
guaranteed. The ground-based community might benefit from pushing
to support the Fermi-LAT mission extension that is currently under
review. Fourty-six contributions have been submitted to {\emph{OG 2.3
    Extragalactic Sources}} (MAGIC: 10, VERITAS: 9, H.E.S.S.: 5,
ARGO-YBJ: 3, Shalon: 3, Fermi-LAT: 2, HAWC: 1, HAGAR: 1, CTA: 2,
Interpretation: 10).

The extragalactic gamma-ray sky is completely dominated by active
galactic nuclei (AGN). All major TeV facilities presented updates on
their searches for extragalactic objects~\cite{BenbowICRC, WagnerICRC,
  BergerICRC, CerrutiICRC}. At TeV energies, 48 extragalactic objects
have by now been discovered. All are identified and all but two are
AGN. These AGN are mostly (38) high-frequency BL LACs (i.e.\ sources
with the jet pointing along the line-of-sight) although several
Flat-spectrum radio quasars (FSRQs) have been discovered by now. 3
radio galaxies have been detected at TeV energies (Cen~A, M~87 and
NGC\,1275). At GeV energies the extragalactic sky ($|b| > 10^{\circ}$
has a significant number of unidentified sources. The only non-AGN
sources detected at TeV energies are NGC~253 and M~82~\cite{OhmICRC},
the most-massive close-by star forming Galaxies that are detected
through their ``galactic diffuse'' gamma-ray emission. Both of these
objects are also detected with the Fermi-LAT. In addition at GeV
energies several local-group galaxies (LMC, SMC, M~31) as well as
other star-forming galaxies have been discovered recently (NGC~4945,
NGC~1068). These detections seem to confirm the relation between
star-formation rate and gamma-ray luminosity, expected in the paradigm
in which cosmic rays are accelerated by supernova remnants or other
objects that are related to star-formation
activity~\cite{FermiStarbursts2}.

Various priors have been tested in the past two years to search for
new extragalactic TeV sources. The most successful ones are a) based
on Fermi-LAT data, using flaring sources at GeV energies to predict
TeV variability~\cite{ErrandoICRC} and also to look for hard-spectrum
sources in the Fermi-LAT data b) optical monitoring campaigns to
predict variability~\cite{BenbowICRC}. Based on these priors, the rate
of detection of extragalactic objects has significantly increased over
the past two years (see Figure~\ref{fig::KifuneExtragalactic}).

In addition to increasing the number of detected objects on the
extragalactic sky, several paradigm shifts have happened over the past
three years. Instead of listing individual new results I would like to
discuss in the following these changes in our study of the
extragalactic sky that are apparent when comparing the results of this
ICRC with earlier results.

\begin{figure*}[!t]
  \centering
  \includegraphics[width=0.8\textwidth]{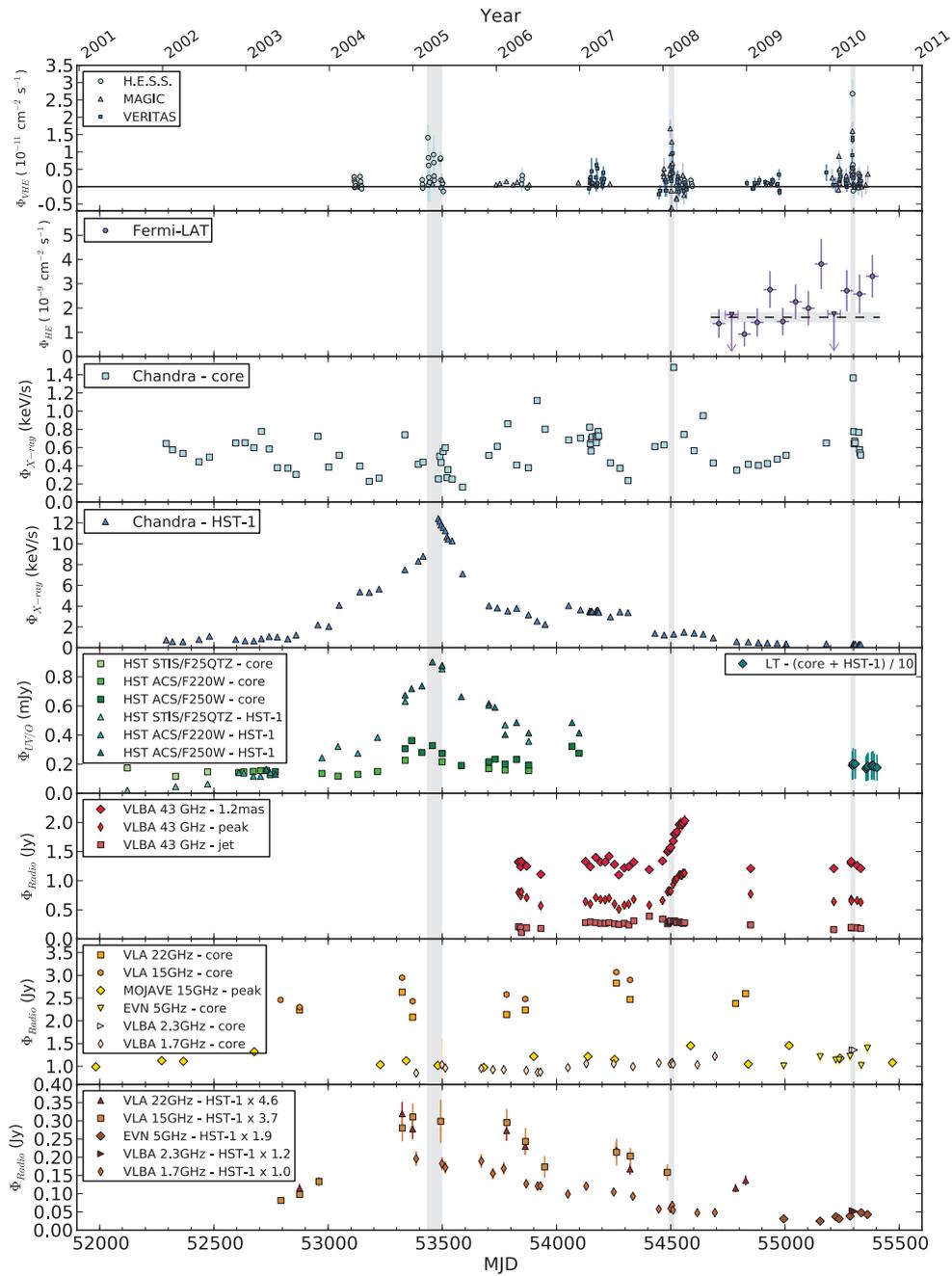}
  \caption{Multi-frequency light curve of M87 from 2001 to 2011. The
    VHE gamma-ray fluxes are calculated above 350 GeV. Gray vertical
    bands show times of increased VHE activity in 2005, 2008 and
    2010. For details, see~\cite{MAGICVERITASFERMIM87}. Reproduced
    from ~\cite{MAGICVERITASFERMIM87}.}
  \label{m87}
\end{figure*}

\begin{itemize}
\item Strictly contemporaneous and broad-band sampling of spectral
  energy distributions (SEDs). As opposed to earlier times when SEDs
  were assembled based on data that was not taken at the same times,
  an effort is now made to obtain data at the same time for different
  wavebands. This is particularly relevant for time-varying sources
  where changes in overall brightness are often accompanied by changes
  in the energy spectra. Given that the Fermi-LAT detects $\sim
  90-95\%$ of all known extragalactic TeV-emitting objects, campaigns
  that involve the Fermi-LAT and several other instruments at higher
  and lower wavelengths have become feasible and are tried to be
  planned strictly contemporaneous. 
\item Related to the first item is that all the TeV instruments are
  involved in massive multi-wavelength campaigns (even
  cross-instrument collaborations at TeV energies). The most
  impressive of those is the multi-year campaign on
  M~87~\cite{GalanteICRC} that involved VERITAS, MAGIC, H.E.S.S. (80
  hours in total), the Fermi-LAT, Chandra, the HST and various other
  instruments all the way down to 1.7 GHz VLBA measurements (see
  Figure~\ref{m87}). 3-4 flaring episodes were detecting in these
  observations that provide a rich data set to study rise-and
  decay-times and spectral properties of flares in this close-by
  non-blazar AGN. Other examples of these large-scale multi-instrument
  efforts include the campaign on Mrk~421 and on
  Mrk~501~\cite{PichelICRC}. A further positive development in this
  regard is that members of the various TeV instruments are now
  conducting regular phone meetings and exchange observation schedules
  to coordinate the searches for extragalactic sources. This is
  particularly valuable as these large collaborations are starting to
  merge into one world-wide CTA community.

\item The TeV instruments are now at a level of about a factor $\sim
  50$ more sensitive than the brightest sources by moving into the
  sub-1\%-Crab regime. Two consequences of this enhanced sensitivity
  are that detailed studies of the historical (and thus bright)
  TeV-emitting blazars (like Mrk421 and Mrk501) can be taken down to
  study 5-minute variability (see e.g.~\cite{PichelICRC}) and that TeV
  instruments are starting to see more than one object in
  extragalactic fields of view (typically $3^{\circ} \times
  3^{\circ}$), see e.g.~\cite{ColinICRC, HildebrandICRC, HolderICRC, BenbowICRC}. In
  hand with this paradigm shift goes that TeV instruments are now
  shifting their observation strategies to move towards higher-quality
  data sets of individual sources rather than trying to simply increase
  the number of sources (VERITAS announced that their observation
  schedule now has 40\% of the time devoted to ``discoveries'' and
  60\% devoted to ``deep observations''). 
\end{itemize}

In terms of modeling the gamma-ray emission~\cite{MankuzhiyilICRC,
  MajumdarICRC, BenbowICRC}, all high-frequency BL-LACs (HBLs) can be
described by sychrotron-self compton (SSC) models, while for the
intermediate-frequency ones (IBLs) SSC combined with external compton
fields are needed to fit the data. For FSRQs the situation is rather
complicated and probably lepto-hadronic models are necessary to fully
explain the data.

Beyond the IACTs with limited fields-of-view continuous monitoring
with non-IACTs is performed. Currently this is only possible for the
brightest TeV sources (such as Mrk421 and Mrk501). ARGO-YBJ reported
of an absence of lags between the X-ray (SWIFT and RXTE), Fermi-LAT
and the TeV emission in flares in these objects~\cite{ZhaICRC}. The
situation will clearly change once HAWC comes online. ~\cite{ImraICRC}
presented sensitivity estimates which showed that the detection of a
factor of 5 increase in flux of Mrk421 will be detected within 1 day
at the 5$\sigma$-level.

Galaxy clusters are the most obvious class of objects where a
detection might be imminent. They are the largest gravitationally
bound structures in the Universe and a reservoir of Cosmic rays is
expected to be present, possibly confined to the cluster over Hubble
times. In addition the annihilation of dark matter might produce a
detectable signal in these objects. So far, no unambiguous firm
detection of a galaxy cluster has been reported above hard X-rays and
limits were presented that put severe constraints on the CR content in
Galaxy clusters. MAGIC spent 80 hours on the Perseus
cluster~\cite{LombardiICRC} and gamma-ray flux limits are now clearly
below predictions from detailed cosmological
simulations~\cite{Pfrommer}. The TeV instruments and the Fermi-LAT
will continue to search for a signal. One would hope to one day combine
the data sets for a given objects (e.g. Perseus) between different
instruments to arrive at even tighter limits (or detections).

Additional highlights of OG 2.3 include:
\begin{itemize}
\item the possible detection of a gravitational lensing-induced
  light-echo from PKS\,1830$-$211 with the Fermi-LAT (but not by the
  LAT collaboration). ~\cite{BarnakaICRC} showed a study of the signal
  following a flaring episode. With a double-power-spectrum method, a
  significance of $4.2\sigma$ is reported for a light-echo flare at
  roughly the right delay time (28 days). However, employing an
  auto-correlation function the significance drops to
  1.1$\sigma$. This system will surely be monitored in the future to
  follow up on this exciting possibility.
\item ~\cite{OrrICRC} put severe constraints on the extragalactic
  background light (EGB) in the infrared by using observations of very
  distant (and hard-spectrum) GeV and TeV AGN.
\item \cite{KachelriessICRC} used the absence of GeV emission for hard TeV
  spectrum AGN (i.e. the absence of a cascading signal) to put
  constraints on the intergalactic B-field to be $>10^{-17}$G.
\end{itemize}

% - Observations of PKS 1222+21 (z=0.43) with MAGIC puts severe constraints on
% emission region due to absence of cutoff and short variability. Only
% the third FSRQ detected in TeV gamma rays (prevalent GeV source class,
% typically at higher redshifts). MAGIC observed it during high GeV
% state. Break at 1-3 GeV, no cutoff in intrinsic (i.e. deabsorbed)
% spectrum implies emission region far from black hole. Flux-doubling
% time: 9min. Means very compact emission region, possibly within the
% large-scale jet. (Becerra-Gonzales- MAGIC)

\section{Gamma-ray Bursts (OG 2.4)}
Also here the Fermi-LAT has had a significant impact on the progress
in the field. It dramatically improved the data at high energies
(compared with EGRET) and now makes more robust predictions for future
instruments such as HAWC and CTA possible. Eleven contributions have
been submitted to {\emph{OG 2.4 Gamma-ray bursts}} (VERITAS: 1, HAWC:
1, ARGO-YBJ: 1, Interpretation: 5, New instruments: 3).

Roughly 682 Fermi-GBM (Gamma-ray Burst Monitor) GRBs have been
detected since August 2008. About half (345) of these occurred within
the field of view of the LAT (defined here at $< 70^{\circ}$). Out of
these 32 LAT GRBs have been detected (9 of which with a recently
implemented new low-energy analysis technique - so-called LLE
events). ~\cite{ConnaughtonICRC} summarized the lessons learned from the
Fermi-LAT: 

\begin{itemize}
\item Extended emission in time (up to 1000s after the initial burst)
  of a delayed high-energy component is detected in several cases
  (e.g. GRB 080916C). The fact that there is a high-energy component
  renders a future detection with ground-based detectors more
  likely. The fact that it is delayed with respect to the initial
  bursts further increases the chance for narrow-field IACTs to detect
  a GRB from the ground in the future.
\item GRB spectra are often not simply described by a so-called
  ``Band function'', but how additional high (and low-) energy
  components (e.g.\ GRB 090902B). 
\item Absorption of the GRB signal on the extra-galactic background
  light (EBL) seems less severe than anticipated in some of the models
  that predicted a large amount of light, meaning that the detection
  of bursts with higher redshifts is possible for a given (high)
  energy.
\item The LAT detects fewer bursts than expected pre-launch when
  predictions were based on a simple extrapolation from BAT-detected
  bursts (at keV energies) to the GeV band. This suggests  breaks in
  the energy spectra between the keV and GeV bands is currently under
  study.
\end{itemize}

  \begin{figure}[!b]
   \vspace{5mm}
   \centering
   \includegraphics[width=0.5\textwidth]{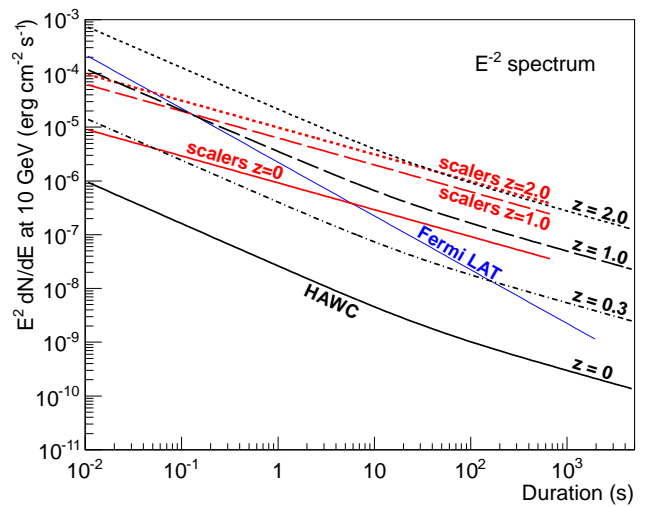}
   \caption{ Sensitivity of HAWC using the main DAQ and the scalers as
     a function of burst duration. The source position was assumed to
     be at a zenith angle of 20$^{\circ}$. The source spectrum was
     assumed as $E^{-2.0}$. Also shown is the flux necessary for the
     observation of 1 photon above 10 GeV by the Fermi-LAT. Reproduced
   from~\cite{HAWCGRBs}. }
   \label{fig::HAWCGRB}
  \end{figure}

Still no ground-based detection ($>50$GeV) has been reported rendering
GRBs the other long-awaited source class (beyond galaxy clusters) that
is not yet detected with TeV instruments. VERITAS reported no
detection in observations of 49 GRBs~\cite{AuneICRC}. It should however be
stressed that the typical delay for the VERITAS observations is $\sim
250$s (the slewing is about $1^{\circ}$/s and there is a human in the
loop). VERITAS is currently investigating whether the slew-speed could
be increased (something MAGIC is already routinely doing, quoting
$<20$s for $180^{\circ}$). All major IACTs have the sensitivity to
detect GRBs, they simply have to be lucky to catch a
not-too-high-redshift GRB within (or close to their) field of
view. The Fermi-LAT team has implemented a GRB-detection algorithm
based on LAT data alone. This will better localize the Fermi-LAT
detected bursts compared to the GBM. Notices are expected to go to the
community within 10s of the GRB.

For the future, the all-sky monitor HAWC promises to become an
extremely sensitive follow-up to MILAGRO, with a factor of 15 improved
sensitivity by 2014. ~\cite{TaboadaICRC} demonstrated that HAWC should
detect bright Fermi-LAT GRBs if the cutoff is $> 60$GeV (the Fermi-LAT
detects events up to $\sim 30$ GeV in e.g. GRB 090902B with no sign of
a cutoff). Additionally, HAWC has received funding for a fourth PMT in
each tank which should decrease the energy threshold further. While
the energy resolution of HAWC at 100~GeV and below will not allow for
a detailed determination of the energy spectrum, the combination of
spectral index and cutoff energy will be constrained using scalars
(see e.g. Figure~\ref{fig::HAWCGRB}). In the more distant future CTA
promises to be a tool to detect GRBs (albeit again with the caveat
that the slewing will take some time). ~\cite{BouvierICRC} presented a
phenomenological model based on Fermi-GBM and LAT measurements where
depending on the way the extrapolation is done about 0.5-1 GRB should
be detectable by CTA per year. If a GRB is detected with CTA there
will be lots of photons detected and the determination of spectral
properties such as energy cutoffs will be performed in exquisite
detail.

\begin{figure*}[!t]
  \centering
  \includegraphics[width=0.85\textwidth]{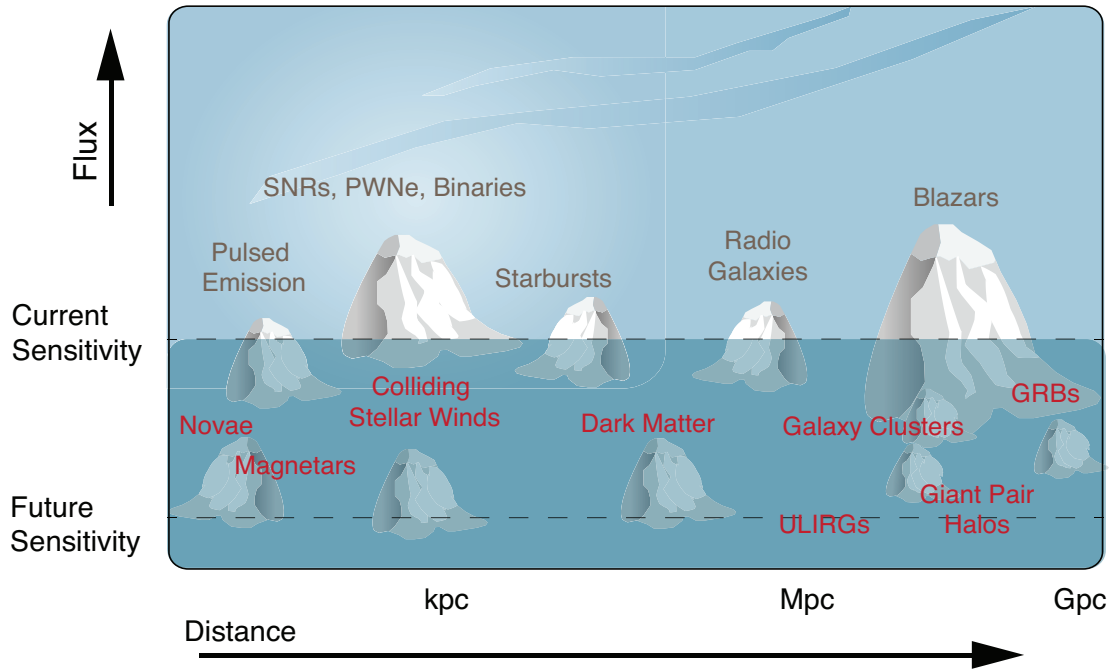}
  \caption{Current status of detection in the TeV gamma-ray
    regime. Adapted from~\cite{HoranWeekes2004}. }
  \label{iceberg}
\end{figure*}

One of the problems that was mentioned is that it is not completely
sure that there will be burst alerts in the time HAWC and CTA
operate. Some of the current missions might end soon or before the
beginning of operation of CTA (e.g.\ SWIFT but
possibly also the Fermi-GBM). SVOM is a Sino-French mission that is
expected to be launched in 2013$-$2014 that will have capabilities
similar to SWIFT. The typical GCN delay expected is 30s for 65\% of
the bursts. UFFO~\cite{ChenICRC} is a revolutionary new concept for
observing early optical photons from GRBs (within 1s) by rotating the
mirrors (instead of rotating the spacecraft). A prototype system will
be launched in 2012 and will provide some burst alerts.

Summarizing session OG 2.4, my prediction is that there will be a
ground-based detection by the next ICRC, either by chance by an IACT 
or by HAWC. 

\section{Instrumentation (OG 2.5)}
A large number of contributions were in session OG 2.5, demonstrating
that the field is actively thinking about future upgrades to existing
instruments and about new developments. Sixty-five contributions were
presented in session {\emph{OG 2.5 Instrumentation}}. CTA alone
presented 26 contributions, HAWC and LHAASO-WCDA each 8, VERITAS 4,
MAGIC and HAGAR 3 and UFFO 4. 9 were ``other'' contributions.

Lots of interesting studies were performed in the past two
years. ~\cite{ZhangICRC} and ~\cite{YaoICRC} reported on studies for LHAASO
(the Large High Altitude Air Shower Observatory), a new planned
detector in the Himalayas which will combine a charged particle array,
a muon-detector array, a wide-field-of-view Cherenkov telescope array
with a Water Cherenkov Array all on one (yet to be specified) site at
high altitude above 5 km~\cite{DanzengluobuICRC}.  In the IACT community,
there was a lot of excitement about the first Geiger-mode APD-based
Cherenkov camera (FACT) which has been installed recently on a
refurbished HEGRA telescope~\cite{KraehenbuehlICRC, VoglerICRC, BretzICRC}. This
will serve as an important test-bed for the application of these
devices in a CTA camera. CTA has provided the largest number of
contributions. It is the one project the whole IACT community (and
beyond) is supporting and pushing for completion. Current plans call
for a start of construction in 2014~\cite{HofmannICRC}. 

In the meantime the currently operating IACTs are all updating their
instruments and are all pushing their thresholds to lower energies to
increase the overlap with the Fermi-LAT. MAGIC has installed a second
17m telescope and is upgrading the camera on the first telescope to
match the properties of the second camera (faster 2 GSamples/s
sampling, increased pixelation from 577 to 1039 pixels and
installation of a sum-trigger). VERITAS has already rearranged their
telescopes to achieve a better geometry for viewing showers and is
installing higher-sensitivity PMTs and a faster trigger for the
camera. H.E.S.S.\ has possibly taken the most dramatic step by
installing a huge 28~m dish in the center of the array. The steel
construction is finished and first light for this telescope is
expected in the summer of 2012.

The potentially most exciting new development is that HAWC is about to
start operation~\cite{GoodmanICRC, ImranICRC, MostafaICRC, BraunICRC,
  HuentemeyerICRC}. ~\cite{BaughmanICRC} showed a first skymap using 16.6
million events from a prototype array with 7 tanks
(VAMOS). Simultaneous all-sky monitoring between the GeV (Fermi-LAT)
and the TeV (HAWC) band will be reported at the next ICRC.

\section{Summary}
The field of gamma-ray astronomy is a very vibrant field with a large
number of talented young researchers demonstrating (and implementing)
exciting new ideas. Significant progress in the understanding of
particle acceleration in the Universe and in the study of the origin
of cosmic rays has been achieved in the past years using instruments
such as AGILE, Fermi-LAT, H.E.S.S, MAGIC and VERITAS.  It is
gratifying to see that the connection between different groups in the
IACT community and also between the IACT and water Cherenkov community
has been growing strongly in the past years. Joint MWL campaigns and
an exchange of observing schedules are routinely done. Going a step
further, shortly after the ICRC IACT collaborations are starting to
exchange data on the Crab Nebula for cross-calibration and also for
the development of common tools in the light of CTA. For all objects
that might be on the verge of a discovery, joining dataset should be
the way forward (e.g. for dwarf spheroidals, Galaxy clusters or for
gamma-ray binaries). As has been learned in the case of dark matter
limits on dwarf spheroidals from Fermi, stacking of dataset can really
improve the significance substantially. My personal highlights were
the presentations on the Fermi-bubbles, the detection of pulsed gamma
rays up to 400 GeV from the Crab pulsar and the flares from the Crab
Nebula.

The field still has lots of open question that we would like to answer
in the future (see Figure~\ref{iceberg}). Where are the Pevatrons?
What is the cosmic ray content in supernova remnants or in Galaxy
clusters? Up to what energies do GRBs accelerate particles? Can giant
pair halos around AGNs be detected? What is the particle nature of
dark matter, is the WIMP scenario valid and if yes, what is the mass
and annihilation cross section of this particle?  Clearly for many of
these questions the scientific output from the community would be
enhanced if the mission of the Fermi-LAT was extended beyond the
currently approved 7 years (i.e. beyond 2015) and if there was an
overlap with HAWC and with CTA to cover sources over as broad an
energy range as possible.

\section*{Acknowledgements}
The author would like to thank the scientific organizers of the
conference for the invitation to give the Rapporteur talk. The local
organizers have been outstanding, in particular Hongbo Hu was a
tremendous help and always ready to help whenever questions or
problems arose. The author would also like to acknowledge the support
by the VERITAS/H.E.S.S./MAGIC collaborations for making the
presentations available beforehand. Also, the author would like to
thank Justin Vandenbroucke for his careful reading of the manuscript.

%  \begin{figure}[!t]
%   \vspace{5mm}
%   \centering
%   \includegraphics[width=2.in]{icrc2011_fig01.eps}
%   \caption{Simple figure example}
%   \label{simp_fig}
%  \end{figure}
% 
%  \begin{figure*}[!t]
%    \centerline{\includegraphics[width=2.in]{icrc2011_fig01.eps}\label{fig2}
%               \hfil
%               \includegraphics[width=2.in]{icrc2011_fig01.eps} \label{fig3}
%              }
%    \caption{An example of a double column floating figure using two subfigures.
%             }
%    \label{double_fig}
%  \end{figure*}
% 
%  \begin{figure*}[th]
%   \centering
%   \includegraphics[width=5in,height=3in]{icrc2011_fig01.eps}
%   \caption{Wide figure example.
%     }
%   \label{wide_fig}
%  \end{figure*}

\end{document}